\documentclass[12pt,preprint]{aastex}

\newcommand{\bvec}[1]{{\mbox{{\boldmath$#1$}}}} 
\newcommand{\eqnref}[1]{(\ref{#1})}

\begin{document}

\title{Subsurface Circulations within Active Regions\footnote{Submitted to the Astrophysical Journal}\,\,$^,$\footnote{Uploaded to arXiv.org on 4/8/09}}

\author{Bradley W. Hindman}
\affil{JILA and Department of Astrophysical and Planetary Sciences,
University of Colorado, Boulder, CO~80309-0440, USA}
\email{hindman@solarz.colorado.edu}

\author{Deborah A. Haber}
\affil{JILA, University of Colorado, Boulder, CO~80309-0440, USA}

\author{Juri Toomre}
\affil{JILA and Department of Astrophysical and Planetary Sciences,
University of Colorado, Boulder, CO~80309-0440, USA}


\begin{abstract} 

Using high-resolution ring analysis (HRRA) we deduce subsurface flows
within magnetic active regions and within quiet sun. With this procedure
we are capable of measuring flows with a horizontal spatial resolution of
2$^\circ$ in heliographic angle (or roughly 20 Mm). From the resulting
flow fields we deduce mean inflow rates into active regions, mean circulation
speeds around active regions, and probability density functions (PDFs)
of properties of the flow field.  These analyses indicate that active regions
have a zonal velocity that exceeds that of quiet sun at the same latitude by
20 m s$^{-1}$, yet active regions advect poleward at the same rate as quiet
sun. We also find that almost all active regions possess a mean inflow
(20--30 m s$^{-1}$) and a cyclonic circulation ($\approx$ 5 m s$^{-1}$)
at their peripheries, whereas their cores, where the sunspots are located,
are zones of strong anticyclonic outflow ($\approx$ 50 m s$^{-1}$). From
the PDFs, we find that active regions modify the structure of convection
with a scale greater than that of supergranulation.  Instead of possessing
an asymmetry between inflows and outflows (with a larger percentage of the
surface occupied by outflows), as is seen in quiet sun, active regions possess
symmetric distributions.

\end{abstract}



\section{Introduction}
\label{sec:introduction}
\setcounter{equation}{0}

The first measurements of the Sun's rotation rate were obtained by careful
tracking of the location of sunspots. In fact, in 1611, shortly after the
invention of the telescope, independent observations of the motion of sunspots
across the solar disk by Galileo Galilei and Christopher Scheiner proved
unequivocally that the Sun rotated. We now know that magnetic features are
not purely passive tracers. Detailed tracking of sunspots and other magnetic
features---such as pores and plages---has revealed that magnetized regions
rotate more quickly than the surrounding field-free plasma
\cite[e.g.,][]{Howard:1970, Golub:1978, Komm:1993}. Furthermore, the larger
the flux concentration the more rapid the rotation rate \citep{Ward:1966,
Howard:1984, Howard:1992}. Recent helioseismic analyses have shown that the
plasma within active regions as a whole also rotates more quickly
\cite[e.g.,][]{Braun:2004}. This superrotation extends to depths as great as
16 Mm \citep{Komm:2009} and there is evidence that the leading polarity might
rotate more rapidly than the trailing polarity \citep{Zhao:2004b, Svanda:2008b}. As in the
earlier studies involving the tracking of sunspots and plage, the helioseismic
studies find that the prograde rotation speed increases with increasing magnetic
flux density.

The tracking of magnetic features has produced less clear results for
meridional motions. We know from direct Doppler velocity measurements
\cite[e.g.,][]{LaBonte:1982, Hathaway:1996} and from helioseismic techniques
\cite[e.g.,][]{Giles:1997, Braun:1998, Basu:1999, Haber:2002, Zhao:2004a}
that within the surface layers, the meridional circulation is poleward
with roughly a speed of 20 m s$^{-1}$. Outside of active regions, correlation
tracking has found that magnetic flux advects towards the pole at roughly
the same speed, acting like a passive tracer \citep{Svanda:2007}. However,
sunspots do not appear to follow the same rules. While individual sunspots
may move substantially in latitude over their lifetime, sunspots as a
group lack systematic poleward motion. Instead, on average, sunspots appear
to slowly drift ($<2$ m s$^{-1}$) away from the center of the active latitude
belts \citep{Woehl:2001, Woehl:2002}.

In addition to their bulk motion, active regions have internal circulations.
As first revealed through the local helioseismic techniques of ring analysis
and time-distance helioseismology, within the surface layers there are large-scale
flows that stream into active regions. These flows typically occupy a layer
7 Mm deep below the photosphere and have amplitudes of 20--30 m s$^{-1}$
\citep{Haber:2001, Gizon:2001}. These same techniques have also demonstrated
that in deeper layers many, but not all, active regions possess strong outflows
with speeds reaching 50 m s$^{-1}$ \citep{Haber:2004, Zhao:2004a}. The largest
magnetic complexes inevitably evince these deep outflows, but many of the
smaller complexes fail to exhibit such behavior. Figure~\ref{fig:DensePack}
shows examples of these organized flows around a large active complex.

This simple circulation pattern (inflows at the surface, coupled to outflows
at depth) becomes more complicated when we consider the flows that are
observed around sunspots. The tracking of ``moving magnetic features" (MMFs)
has revealed an annular collar of outflow surrounding sunspots
\cite[for a review see ][]{Hagenaar:2005}. This ``moat" of flow has also been
observed helioseismically \citep{Lindsey:1996, Gizon:2000, Braun:2003}. Related
outflows have also been observed surrounding newly emerged active regions that
have yet to fray and form a region of extended plage \citep{Hindman:2006b,
Komm:2008, Kosovichev:2008}. Figure~\ref{fig:EmergingAR} shows the flows around
an active region that emerged just two days prior. The flow field during this
young stage in the active region's life is dominated by strong outflows from
the sunspots.

In this paper we perform detailed measurements of the flows in quiet sun
and in active regions using the local helioseismic technique of ring analysis
to detect flows within the upper 2 Mm of the solar convection zone. Our goal
is to understand the circulations that are typically established within active
regions and to assess how properties of the flow field---such as its divergence
and vorticity---differ between magnetized and nonmagnetized regions. In
\S\ref{sec:HRRA} we discuss the ring analysis procedure in detail. In \S\ref{sec:ARFlows}
we present two different schemes for analyzing the measured flow fields.
In \S\ref{sec:Discussion} we provide a comprehensive discussion of our findings,
and in \S\ref{sec:Conclusions} expound our conclusions.


\section{High-Resolution Ring Analysis (HRRA)}
\label{sec:HRRA}
\setcounter{equation}{0}

Ring analysis assesses the speed and direction of subsurface horizontal
flows by measuring the advection of ambient waves by the flow field. In
the presence of a flow, waves traveling in opposite directions have their
frequencies split by the Doppler effect, providing a direct measure of the
flow velocity in those layers where the waves have significant amplitude.
For this particular study, we have utilized only surface gravity waves
($f$ modes); therefore, we are able to probe a layer several Mm thick lying
directly below the photosphere. The frequency perturbation introduced by
the flow is $\Delta\omega = \bvec{k} \cdot \bvec{\hat{U}}$, where $\bvec{k}$
is the horizontal wavenumber and $\bvec{\hat{U}}$ is the integral over depth
of the horizontal flow velocity weighted by a kernel which is approximately
the kinetic energy density of the surface gravity wave.

The frequency splittings produced by flows are measured in the Fourier domain.
For a single analysis, a power spectrum is obtained of the wave field in a
localized region on the solar surface by Fourier transforms (two in space,
one in time) of a sequence of tracked, remapped, and apodized Dopplergrams
\citep{Bogart:1995, Haber:1998}. The mode power in the spectrum is distributed
along curved surfaces, which when cut at constant frequency appear as a set
of concentric rings, each corresponding to a mode of different radial order.
These rings are nearly circular in shape with centers displaced slightly from
the origin due to the splitting of the mode frequencies. The frequency splittings
$\Delta\omega$ are obtained as a function of wavenumber by carefully fitting
the $f$ modes in such power spectra with Lorentzian profiles
\cite[e.g.,][]{Haber:2000, Haber:2001}.

A single ring analysis is performed within a small region, or tile, on the
solar surface. The resulting flow is an average measure of the flow within
that tile \citep{Hindman:2005}. We map the flow field over the entire visible
disk by performing many such ring analyses on a mosaic of different locations
on the solar surface. Each day the analyses are repeated to obtain a sequence
of daily flow maps.

For this study we have used Dynamics Program data from MDI \citep{Scherrer:1995}
on the {\sl SOHO} spacecraft. We use the same tiling and tracking scheme discussed
at length in \cite{Hindman:2006a}. This scheme, known as high-resolution ring
analysis (HRRA), produces a daily map of the flow field formed from the analysis
of over $10^4$ tiles that are 2$^\circ$ in heliographic angle on a side. The
tiles overlap and their centers are separated by $0^\circ.9375$ in longitude and
latitude. In order to reduce the amount of image tracking and remapping that
is required, in practice, instead of tracking each of these tiles separately,
we track 189 larger regions that span 16$^\circ$ in angle on a side \citep{Haber:2002}.
The smaller 2$^\circ$ tiles are then extracted from these larger tracked regions.
Each of the larger tiles, is tracked for 27.7 hours at the surface rotation rate
appropriate for the center of the tile \citep{Snodgrass:1984}. The large tiles
overlap each other and their centers are separated in longitude and latitude by
7.5$^\circ$.

Since the large tiles overlap, for any given location in the mosaic of 2$^\circ$
tiles, there exists multiple realizations of that tile each extracted from a different
neighboring large tile. In general, any given location will have 4 separate tiles
and the flow determinations from these different realizations are averaged together.
Since, not all of the neighboring large tiles are tracked at the same rotation rate,
before averaging we must convert each of the flow realizations to a common rotation
rate.  This is accomplished by subtracting from the zonal velocity a longitudinal
and temporal mean obtained from the small tiles from a given year. Therefore, the
reported zonal flows are measured relative to the differential rotation rate obtained
from the data itself.

As in \cite{Hindman:2006a}, the results for waves of different horizontal wavenumber
are averaged together to increase the ratio of signal to noise. A typical daily
flow map, comprised of roughly 10$^4$ measurements, is generated from $1.3\times 10^5$
separate flow determinations. Assuming that the errors are uncorrelated, this
averaging procedure produces a fractional uncertainty of roughly 20\% for any
single flow speed measurement in the daily map. Figures~\ref{fig:EmergingAR}
and \ref{fig:LargeAR} show flow maps obtained by this technique. Figure~\ref{fig:EmergingAR}
shows the flows around a newly emerged active region where the dominant flow
structure is an outflow from the sunspots. Figure~\ref{fig:LargeAR} shows the
flows near a large, mature active complex, which has persisted for
several solar rotations and undergone multiple flux emergence events.

The spatial resolution achieved by this technique is largely determined
by the size of the analysis tiles---which determines the wavelength of the
waves that are sampled; however, the shape of the apodization function, the
details of the fitting procedure and the damping length of the waves may
also play a role \citep{Hindman:2005, Birch:2007}. For the tiles used in
this study, we expect that the shape of the averaging kernel is essentially
the product of a vertical profile and a horizontal planform. The vertical
profile is provided by the standard $f$-mode eigenfunction and the horizontal
planform is a singly-humped function that vanishes at the edge of the tile
and peaks at the tile center \citep{Birch:2007}. Since the $f$ modes are
surface gravity waves, they are confined to a narrow layer just below the
solar surface. The exact depth to which the eigenfunction extends is proportional
to the horizontal wavelength. However, the small tiles used in this study
permit only a relatively narrow band of wavelengths to be measured. Thus,
the vertical eigenfunctions of the measured $f$ modes are rather similar
and the flow measurements are essentially a mean over a layer spanning the
first 2 Mm below the photosphere.

\section{Flows within Active Regions}
\label{sec:ARFlows}
\setcounter{equation}{0}

We have generated daily flow maps using MDI Dynamics Program data for three
periods of time in three subsequent years: 1 March to 26 May 2001, 11 January
to 21 May 2002 and 12 September to 15 November 2003. In total, due to gaps
in some of these periods, we have produced flow maps for 201 days of data.
We have analyzed the measured flow maps with two distinct procedures. The
first is the calculation of probability density functions (PDFs) for a variety
of flow parameters within both quiet sun and within regions of magnetism.
From these PDFs we examine how the mean properties of the flow vary with
magnetic activity as well as how the shapes of the distributions change. The
second analysis involves identifying active regions, measuring spatially
structured flows within those regions and computing average active region
flow structures.

\subsection{Probability Density Functions for Flow Properties}
\label{subsec:Distributions}
\setcounter{equation}{0}

We compute distribution functions for the zonal and meridional components
of the flow field as well as for the divergence and vorticity of the flow. Since
our HRRA procedure only produces estimates of the horizontal flow, the
divergence that we compute is only the horizontal divergence and the vorticity
is only the vertical component of the vorticity. Since we are interested in the
differences in the flow field between magnetized regions and quiet sun, we
need to analyze active pixels separately from quiet pixels. This requires
that we produce colocal flow maps and magnetograms with the same pixel spacing.
We achieve this by using MDI magnetograms and interpolating our HRRA flow maps
to the same spatial sampling via splines. Simultaneously, we compute the spatial
derivatives necessary for the divergence and the vorticity by utilizing the same
spline coefficients. PDFs are computed for quiet sun through the selection
of flow measurements from only those pixels with a field strength less than
50 G. Separately, PDFs for magnetized regions are calculated using pixels
with a field strength greater than 50 G. Furthermore, since we expect that
many of the flow properties may be functions of latitude, we compute separate
PDFs for different latitudinal bands. There are 11 bands in total, each
$10^\circ$ wide, with centers separated by 10$^\circ$ and evenly distributed
about the equator. Each quiet sun PDF is constructed of over 10$^5$ independent
flow measurements, whereas, due to their low filling factor, the PDFs for active
regions are composed of roughly 10$^4$ distinct measurements. As a test, we have
varied the width and number of latitudal bands, finding little difference in
the results except for the expected changes in error estimates due to the number
of data points in each band. For simplicity, we have chosen to show only the
results for the 10$^\circ$ bands.

Figures~\ref{fig:ZonMerPDFs} and \ref{fig:DivCurlPDFs} present the distribution
functions for active and quiet regions. For clarity, we have only shown
the distributions for the latitudinal bands centered at $30^\circ$ north and
south of the equator. Figure~\ref{fig:MeanFlows} shows the mean values of
these distributions as a function of latitude. From this set of figures one
can clearly see that, on average, magnetized regions rotate across the solar
disk more rapidly than quiet regions (by roughly 20 m s$^{-1}$), yet magnetized
regions appear to move poleward at the same rate as quiet regions.  Furthermore,
the meridional circulation is poleward, with a nearly sinusoidal shape as a
function of latitude. We see no evidence for residual circulations in the
active bands as seen by some helioseismic studies \cite{Zhao:2004a, Gonzalez-Hernandez:2008}.
this may be the result of averaging the flows over three separate years and the
relatively small size of these residual circulations ($\approx$ 5 m s$^{-1}$).
The distributions of the divergence reveal that active regions appear as zones
of converging flow, while
quiet sun is on average slightly divergent. Finally, magnetized regions possess
cyclonic vortical motions that increase linearly with latitude. These findings
will be discussed in more detail in \S\ref{sec:Discussion}.

In order to examine the shape of the distributions with a better signal-to-noise
ratio, we average the distributions over latitude. Since the PDFs for the
zonal flow and divergence are largely independent of latitude, we average
these over all latitudes. The meridional flow and the vorticity are antisymmetric
with latitude; therefore, we average those separately over each hemisphere.
The results are shown in Figure~\ref{fig:MeanPDFs}. The mean shifts between
magnetized and quiet regions are obvious in these figures. However, it is now
also  clear that the shape of the distributions change within active regions.
In particular, the distributions of all flow quantities within active and quiet
regions have similar cores, but those in active regions possess extended wings,
indicating that a wider range of speeds is present. Furthermore, in the
divergence distributions, the flows in the quiet sun are notably asymmetric,
with more area occupied by diverging flows than converging flows. On the other
hand, in active regions, the flows are quite symmetric. Typical widths for the
zonal and meridional flow distributions are 80 m s$^{-1}$, while the divergence
and vorticity have widths on the order of 20 $\mu$Hz and 10 $\mu$Hz, respectively.

\bigskip
\centerline{TABLE 1}
\centerline{\small\scshape Moments of the Distributions}
\begin{center}
{\footnotesize
\begin{tabular}{lccccc}
\hline\hline
{} & {} & \multicolumn{2}{c}{Quiet Regions} & \multicolumn{2}{c}{Magnetized Regions}\\
{Flow Property} & {Hemisphere} & {Mean$^1$} & {Variance$^1$} & {Mean$^1$} & {Variance$^1$}\\
\hline
Zonal Flow 	& Both 	& -1.1 	& 81.6 	& 19.0 	& 87.8 \\
Meridional Flow & North & 14.0 	& 82.0 	& 10.0 	& 94.6 \\
Meridional Flow & South & -20.4 & 82.0 	& -19.8 & 79.2 \\
Divergence 	& Both 	& 0.2 	& 17.0 	& -4.4 	& 17.1 \\
Curl		& North & 0.005	& 9.6 	& 0.4 	& 15.7 \\
Curl		& South & 0.01	& 8.9 	& -0.7 	& 12.1 \\
\hline

\multicolumn{6}{l}{$^1$The means and variances for the zonal and meridional flow are measured in}\\
\multicolumn{6}{l}{~~units of m s$^{-1}$. The divergence and vorticity are measured in units of $\mu$Hz.}\\
\multicolumn{6}{l}{~~The variance is defined, as usual, as the second central moment of the distribution.}
\end{tabular}

}
\end{center}

\subsection{Flow Structures within Active Regions}
\label{subsec:Structures}
\setcounter{equation}{0}

In order to assess the importance of organized flow structures within active
regions, we have chosen to average flow properties over different zones within
an active region. These zones are identified by various contour levels in
smoothed maps of the unsigned magnetic flux. Figure~\ref{fig:MagGram} shows
an MDI magnetogram of NOAA AR9433 (the same active region shown in
Figure~\ref{fig:LargeAR}). The overlying contours were obtained by smoothing
the modulus of the magnetogram with a Gaussian filter with a width of $2^\circ$
in heliographic angle---the same spatial resolution as the helioseismic flow
measurements. Any region with magnetic field strength greater than 50 G was
labeled active. Only regions within $60^\circ$ of disk center were considered
since the helioseismic measurements have a similar coverage. Figure~\ref{fig:DistARs}
shows distributions of the area and total magnetic flux of the resulting set
of flux concentrations. We further winnowed our sample of flux concentrations
by selecting only the subset with areas greater than $1\times10^4$ Mm$^2$. Over
the course of the 201 days of data, over 100 independent flux concentrations
meeting these criteria were identified.

The flows coincident with each of the magnetic contours within each of these
active regions are decomposed into an inflow component (perpendicular to the
contour and pointed inwards) and a circulation component (parallel to the
contour and pointed counterclockwise). For each active region, a line integral
was performed around each contour to compute the mean inflow speed and the
mean circulation speed. The results for each strength of magnetic contour
were then averaged over all active regions in the sample to form mean inflow and
circulation speeds as a function of the magnetic field strength $B$.

\begin{eqnarray}
v_{\rm inflow}(B) &\equiv& \sum_i \frac{w_i(B)}{L_i(B)}
		\int_{C_i(B)} \bvec{v} \cdot \bvec{\hat{n}} \; dl\; ,	\\	
\nonumber \\
v_{\rm circ}(B) &\equiv& \sum_i \frac{w_i(B)}{L_i(B)}
		\int_{C_i(B)} \bvec{v} \cdot \bvec{\hat{t}} \; dl\; ,	\\
\nonumber \\
\nonumber L_i(B) \equiv \int_{C_i(B)}  dl\; , &&\;\;\; w_i(B) \equiv \frac{L_i(B)}{\sum_j L_j(B)}\; .
\end{eqnarray}

In the expressions above, the integrals are contour integrals around the
magnetic contour $C_i$ with magnetic flux density $B$ of the $i$th active region
in our sample. The unit vectors $\bvec{\hat{n}}$ and $\bvec{\hat{t}}$ are,
respectively, normal and tangential to the magnetic contour, with the normal
pointing inward toward higher field strength and the tangent pointing
counterclockwise around the contour. Each integral is divided by the length
of the contour $L_i$ to generate an average speed and the summation is a
weighted average over all active regions, with the weights, $w_i$, proportional
to the length of the contour.
 
The above averaging process was performed separately for active regions whose flux
weighted centers lie in the northern hemisphere and those in the southern.
The results are shown in Figure~\ref{fig:ARflows}. The periphery of active
regions possess positive inflows with an amplitude of 20--30 m s$^{-1}$ and
have a weak tendency for cyclonic circulation with a speed of 5 m s$^{-1}$.
As the field strength $B$ increases towards the interior of active regions,
this inflow turns into an outflow, presumably forming a downflow where
the two flows meet. The very core of active regions, formed by sunspots, are
zones of strong outflow ($\approx 50$ m s$^{-1}$) and anticyclonic motion
with a rotational speed of approximately 10 m s$^{-1}$.


\section{Discussion}
\label{sec:Discussion}
\setcounter{equation}{0}

Using HRRA we have estimated the horizontal flow field within the upper 2 Mm
of the solar convection zone by measuring the Doppler shifts of $f$ modes.
Flow maps for 201 days of data with a horizontal resolution of 2$^\circ$ have
been produced and two separate analysis procedures applied. Firstly, from
these flow maps we have calculated PDFs of the zonal flow, the meridional
flow, the flow's divergence and it's vorticity. Secondly, we have computed mean
inflow rates and circulation speeds at various magnetic field strength levels
within active regions. From these analyses we deduce the following, and will
expand upon these findings in the subsequent subsections.

\begin{itemize}
	\item Magnetized regions rotate more rapidly across the solar disk than
	nonmagnetized regions (by roughly 20 m s$^{-1}$).

	\item Magnetized regions are advected poleward by the meridional circulation
	at the same rate as quiet regions.

	\item On average, magnetized regions possess convergent cyclonic vortical
	motions, whereas the flows in quiet sun are weakly divergent without
	a measurable vortical preference.

	\item The flows within active regions span a wider range of flow speeds
	than those seen in quiet sun.

	\item The divergence distribution in quiet sun peaks at zero, but is
	asymmetric about this peak value, with more of the solar surface covered
	by outflows than inflows. The divergence distribution in magnetized
	regions	peaks at a negative value (converging flows), and is symmetric
	about its peak.

	\item The periphery of an active region is a zone of inflow (20--30 m s$^{-1}$)
	as well as a zone of cyclonic circulation ($\approx 5$ m s$^{-1}$).

	\item The moat flows streaming from sunspots form anticylones with a
	mean rotational speed of roughly 10 m s$^{-1}$.
\end{itemize}

\subsection{Bulk Motion of Active Regions}
\label{subsec:BulkMotions}

We find that active regions, on average, rotate across the solar surface
with a speed that is roughly 20 m s$^{-1}$ faster than quiet sun at the same
latitude. Our observation that active regions are zones of superrotation
is consistent with the previous findings obtained through feature tracking,
direct Doppler velocity measurement and helioseismology. The rate of
superrotation (20 m s$^{-1}$) also agrees with previous findings if one
takes into consideration the spatial resolution of the various measurement
schemes. Surface tracking of magnetic features \cite[for a review see ][]{Howard:1996}
and high-resolution helioseismic measurements \citep{Braun:2004, Zhao:2004b}
indicate that sunspots rotate at a rate that is roughly 50 m s$^{-1}$ greater
than quiet sun, whereas the rotation rate of plages is markedly less. The
low-resolution ring-analysis procedure employed by \cite{Komm:2008} finds
that active regions as a whole superrotate at a rate of 4 m s$^{-1}$. This
low-resolution procedure has a spatial resolution of 15$^\circ$, and is thus
incapable of resolving sunspots. In fact, a single flow measurement averages
over sunspots, surrounding plage and even a significant portion of quiet sun.
Therefore, one would expect a dilution of the superrotation rate. To a lesser
degree the same averaging effect occurs here. Our horizontal spatial resolution
is 2$^\circ$, which is sufficient to resolve an active region, but still
incapable of resolving sunspots. Therefore, we would expect that our
superrotation rate should lie between that for sunspots and that for plage.

The meridional component of the flow field within active regions behaves
rather differently than the zonal component. Instead of moving at a rate
that differs from quiet sun, we find that the fluid within active regions
advects poleward at exactly the same speed as quiet sun. While this result
confirms previous surface measurements \cite[e.g.,][]{Svanda:2007}, this
is the first time that such a result has been reported for helioseismic
measurements. It is not entirely clear how this evident poleward advection
of the bulk of the active region can be reconciled with the observation that
sunspots lack systematic poleward motion \cite[e.g.,][]{Woehl:2001, Woehl:2002}.

When a rising flux rope first emerges through the solar surface, we expect
that the magnetized region will be marked by an upwelling that swells
horizontally---due to the decreasing gas pressure with height in the solar
atmosphere. At this point in the active region's evolution, the magnetic
field is dynamically important both at the surface and below. Thus, the
field at the solar surface is influenced by the flows all along the flux
rope. This may explain why magnetic regions rotate more quickly than the
quiet sun. The Sun's rotation rate increases with depth throughout the
upper 30 Mm of the convection zone and the magnetic field in the sunspot
might be grabbed and dragged by the fast moving subsurface layers
\cite[e.g.,][]{Gilman:1979, Hiremath:2002, Sivaraman:2003}. However, if
similar arguments are made for meridional advection of magnetic elements
a contradiction arises. Plage within active regions and small magnetic
elements outside of active regions are observed to passively advect with
the meridional circulation. This observation is consistent with the
helioseismic determinations of the meridional flow, since those measurements
indicate that the flow remains roughly constant with depth in the upper 20 Mm
of the convection zone \citep{Haber:2002, Zhao:2004a}. However, the lack of
systematic meridional motion by sunspots, by the same arguments, would appear
to indicate that the sunspots must be rooted very deeply, within the supposed
return flow in the meridional circulation. This is clearly in contradiction
with the helioseismic observations that fail to find such a return flow within
the near-surface shear layer, which is where we have presumed that the sunspots
are anchored in order to explain their superrotation.

This conundrum becomes even more complicated if we consider what might
happen as an active region ages. \cite{Fan:1994} have suggested that
the connection between the field at the surface and its underlying roots
may be broken through a dynamical disassociation process. Their original
suggestion involved the establishment of hydrostatic equilibrium throughout
the length of the tube once it emerges. Due to differences in entropy
stratification between the fluid within the tube and the external field-free
gas, at a layer roughly 10 Mm below the solar surface, the internal and
external gas pressure match. Therefore, lateral pressure balance requires
that the magnetic pressure vanish, the tube herniates and turbulent
convection shreds the tube because the magnetic field becomes dynamically
insignificant. The field above the herniation layer becomes disconnected
from the field below, thus enabling the surface field to be advected and
dispersed by near-surface flows, a process that is presently well-modelled
by surface transport models \cite[e.g.,][]{Wang:1989, van Ballegooijen:1998,
Schrijver:2001, Baumann:2004}.

\cite{Schuessler:2005} have pointed out that the establishment of hydrostatic
equilibrium along the entire length of the flux rope is much too slow
a process. Observations of the changing dynamics of magnetic structures in
the photosphere indicate that the field begins to disconnect a couple of days
after emergence. \cite{Schuessler:2005} have refined the dynamical disconnection
model by demonstrating that surface cooling in regions of intense magnetism
can drive downflows that both concentrate the field at the surface (through
evacuation and collapse) and enhance the dynamical disconnection at depth
by increasing the subsurface pressure where the deep upflow along the flux
rope and the surface driven downflow meet. This process operates quickly,
on a time scale of several days after emergence, and recent helioseismic
studies indicate that the flows within active complexes change from upflows
to downflows over such a period of time \citep{Komm:2008}. The mechanism may
explain the observation that young sunspots rotate more rapidly than old
sunspots \cite[e.g.,][]{Balthasar:1982, Svanda:2008a}. Initially, the sunspot is dragged
by rapidly rotating subsurface layers, but after disconnection the sunspot
slows due to viscous or turbulent drag. Similarly, the observed advection
of plage and weak field both inside and outside active regions is
well-explained. Dynamical disconnection allows the magnetic flux to advect
like a passive tracer since its roots have been severed. We once again find
that the sticking point is the lack of systematic meridional motion by sunspots.
If sunspots become disconnected from their roots in a matter of 2 or 3 days,
we would expect that the meridional circulation would exert its influence and
begin to advect the sunspots poleward.

\subsection{Steady Flows Established around Mature Active Regions}
\label{subsec:SteadyFlows}

In addition to the bulk motions of active regions, we have measured internal
flow structures. Two separate but related measurements (the mean divergence
and the mean inflow speed) show that within the surface layers, active regions
are zones of convergence. From the PDFs of the divergence we see that on average,
magnetized regions possess a negative divergence (see Figures~\ref{fig:MeanFlows}
and \ref{fig:MeanPDFs}). Note however, that the variance about this mean value
for the divergence is much larger than the mean itself (see Table 1).  Equivalently,
our measurements of the mean inflow speeds within the surface layers of active
regions show that almost all active regions possess a net inflow of 20--30 m s$^{-1}$
at their periphery. Conversely, the cores of active regions, formed by the presence
of sunspots, possess strong outflows. These outflows arise from the moat
flows that extend 10--20 Mm beyond the penumbra \cite[e.g.,][]{Sheeley:1972,
Harvey:1973, Gizon:2000, Hagenaar:2005}.  The presence of inflows at the
periphery and outflows from sunspots dictates that somewhere within the
active region's plage, the two flows meet and downflow must occur.
Presumably, the downflow within the plage connects to the deep outflows
that are seen to emerge from active regions at depths greater than 10 Mm
in low-resolution helioseismic flow measurements \citep{Haber:2003,
Haber:2004}. This downflow may also partially supply the return flow
needed for the moat flow observed at the surface, although as of yet
helioseismology has not been able to tell us how deeply this return
flow may be rooted.  Figure~\ref{fig:ARsideview} shows a cartoon sketch
of the inferred circulations. The flows indicated with the arrows outlined
in white have been directly observed through helioseismic techniques
\cite[e.g.,][]{Gizon:2000, Braun:2003, Haber:2003, Gizon:2005} as well
as through direct Dopper velocity measurements \cite[e.g.,][]{Sheeley:1972}
and magnetic feature tracking \cite[e.g.,][]{Brickhouse:1988}. The remaining
flows are a logical means of connecting the observed components
of the flow field; however, without direct helioseismic measurement of the
vertical component of the flow---which still remains elusive, the topology
of this circulation remains speculative. 

We also find systematic circulations around active regions. From the PDFs
of the vertical component of the vorticity (see Figures~\ref{fig:MeanFlows}
and \ref{fig:MeanPDFs}) we deduce that magnetized regions have a tendency
to possess cyclonic vorticity. Consistently, we also measure a mean cyclonic
circulation speed of roughly 5 m s$^{-1}$ around the peripheries of active
regions. Just as the inflows transition to outflows as we move from the edge
of an active region toward the sunspots, the circulations transition from
cyclonic flows around the boundary to anticyclones at the location of the
sunspots.

\subsection{Source of the Inflows and Circulations}
\label{subsec:Source}

One possible mechanism for the generation of the surface inflows into active
regions has already been mentioned. Plage and faculae are bright and, therefore,
locations of radiative cooling \cite[e.g.,][]{Fontenla:2006}. Radiative cooling
in magnetized surface layers will generate downflows as cool, dense material
looses buoyancy and plunges into the solar interior. Such downflows draw
fluid from the surrounding surface layers, generating inflows into
magnetized regions. The effect of enhanced surface cooling may be further
augmented by a reduction in the convective energy flux within regions of
magnetism, arising from the systematic suppression and modification of granulation.

One consequence of such a model is that the inflow speed may well be a function
of the area occupied by the active region's plage. The radiative cooling and
hence the mass downflow rate should be proportional to the area of the plage,
whereas the mass flux into the active region is proportional to the inflow
speed and the circumference of the active region. If the mass supply rate
from the inflow is to equal the subduction rate, the inflow speed must increase
as the square root of the area of the plage. Furthermore, the inflow should
extend for some distance into the quiet sun around the active region. To date,
no attempt has been made to detect correlations between inflow speed and
active region size nor has the inflow been systematically measured in quiet
sun in the vicinity of active complexes. We plan to pursue such studies
in the near future.

If the surface cooling mechanism is correct, the moat flows
that stream from sunspots would seem to indicate that sunspots do not participate
in the same radiative cooling, or perhaps the surface cooling within sunspots
is significantly weaker than it is in plage. However, the facts that moat flows
do not appear isotropically around all sunspots, and that the presence of moat
flows appears to be connected to the existence of penumbra \citep{Vargas Dominguez:2007,
Vargas Dominguez:2008}, provide evidence that another mechanism may be at work.
Depending on the mechanism, the depth of the return flow is likely to be very
different. Using time-distance helioseismology with $p$ modes (as opposed to $f$ modes),
\cite{Zhao:2001} have reported the observation of a returning inflow around a
sunspot that spans a depth range of 1.5--5 Mm. However, the connection that these
inflows may have to the moat flows is difficult to assess, since the moat flows
themselves are not detected within that study.  

The circulations that are established around active regions seem to be well
correlated spatially with the observed inflows and outflows. In the region
where we observe inflows into active regions, we measure cyclonic motion
rotating around the active region. In those regions with outflows, we also
measure anticyclones. We suspect that this correlation is not accidental.
A likely explanation is that the circulations are caused by the deflection
of the inflows and outflows by the Coriolis force. Therefore, the ultimate
source of the circulations is the same mechanism that drives the inflows
at the periphery and the mechanism that produces the moat flow around
sunspots. If we assume that the flows are steady, and that the inflows are
driven by a pressure gradient, we may estimate the size of the circulation
speed from the momentum equation in a rotating coordinate system,

\begin{equation}
(\bvec{v}\cdot\bvec{\nabla})\bvec{v} = -\frac{1}{\rho}\bvec{\nabla}P - 2\bvec{\Omega}\times\bvec{v}\; ,
\label{eqn:Momentum}
\end{equation}

\noindent where we have explicitly dropped the partial derivative with respect
to time in the advective derivative because the flows are steady. The component
of this equation tangential to the pressure gradient is a balance purely
between the advective derivative and the Coriolis force. For simplicity,
we employ an $f$-plane approximation and consider an active region that is
circular in shape. In polar coordinates $(r,\phi)$ with the origin at the active
region's center, the circulation speed, $v_{\rm circ} = v_\phi$, is given by
the angular component of equation~\eqnref{eqn:Momentum}

\begin{equation}
v_r \frac{\partial v_\phi}{\partial r} = -2\Omega v_r \sin\theta \; .
\end{equation}

\noindent If we assume that the flow extends over a radial distance $H$,
then the circulation speed may be estimated by

\begin{equation}
v_{\rm circ} \sim 2 \Omega H \sin\theta \; .
\end{equation}

If we assume that $H$ equals the radius of a typical active region, 100 Mm,
and we use the Carrington rotation rate, $\Omega = 456$ nHz, we estimate a circulation
speed of 46 m s$^{-1}$ at a latitude of 30$^\circ$. Clearly, this estimate
depends critically on $H$, the distance over which the inflow extends into quiet
sun. But for reasonable values, the Coriolis force is strong enough to generate
circulation speeds on the order of 5 m s$^{-1}$. Furthermore, the effect should
increase with latitude and should be antisymmetric about the equator, as
is reproduced in Figure~\ref{fig:MeanFlows}.

A similar mechanism was suggested by \cite{Spruit:2003} as the source of
the torsional oscillations. In his model, radiative cooling is enhanced in
the active region belts by reduced opacity within the small-scale magnetic
fibrils forming the plage \cite[e.g.,][]{Spruit:1977}. This induces a slight
decrease in temperature and pressure within the active latitudes. Coriolis
forces generate steady geostrophic flows around these low pressure regions
and the torsional oscillations are simply the resulting thermal wind. One
possible extension of Spruit's model would be to argue that geostrophic
balance is established around each active region separately as the radiative
cooling is localized to the plage. A longitudinal average of these geostrophic
circulations would result in a zone of faster rotation at the low-latitude
edge of the active region belt and a slow down at the high-latitude edge.

In our case, we measure circulations around each active region that possess
the same handedness as the geostrophic flows in Spruit's model and the measured
circulation speed ($\approx 5$ m s$^{-1}$) is comparable to the amplitude of
the torsional oscillations, it is reasonable to ask if our circulations are
the source of the torsional oscillations. If this were so, the amplitude of
the torsional oscillation should be the mean active region circulation speed
multiplied by the fraction of longitudes occupied by magnetic activity. Since
the circulation speed measured here is only 5 m s$^{-1}$, a dilution by a
longitudinal filling factor would result in torsional oscillations with an
amplitude that is only a fraction of the observed value. Of course the extent
to which the flows protrude into quiet sun will reduce the dilution factor.
We should also point out that the torsional oscillation pattern extends to
high latitude during the quiet phase of the solar cycle when active
regions are largely absent \citep{Schou:1999, Basu:2003, Howe:2005, Howe:2006}.
This property suggests that another mechanism is at work either in isolation
or in conjunction with active region circulations.

\subsection{Convective Motions of Active Regions}
\label{subsec:Convection}

It has long been known that granulation is modified and perhaps suppressed
within regions of intense magnetism. Whether this suppression applies to
larger scales of convection is not as clear. We find here that for scales of
motion larger than supergranulation, the flows appear to be less organized
within magnetic active regions; they lack the regular tiling of convection
cells that is apparent within quiet sun (see Figure~\ref{fig:LargeAR}).
Despite the disruption of the cellular pattern, the flows speeds within active
regions are generally larger than in quiet sun. This increase in speeds isn't
caused by a general broadening of the distribution. Instead the PDFs
evince elevated tails for flow speeds in excess of 200 m s$^{-1}$.

Another noticeable difference between the PDFs in active and quiet regions
is the shape of the divergence distribution. In quiet sun, the distribution
is asymmetric with an enhanced wing corresponding to positive values of the
divergence. Therefore, a larger percentage of the solar surface is occupied
by divergent outflows than convergent inflows. This property is consistent
with the asymmetric nature of solar convection that is seen in numerical
simulations of solar convection, where the convection is composed of zones of
broad, upwelling outflows in the center of convection cells and the narrow,
inflowing downflows at the cell boundaries \cite[e.g.,][]{Stein:1989, Stein:1998,
Cattaneo:1991, Brummell:1996}. This asymmetry arises from the gravitational
stratification.  Upflows expand as they rise because of the decreasing gas
pressure. Downflows contract into plumes as they descend for the same reason.
Note that the measured PDFs for the divergence do not exhibit the bimodal structure
seen in large-scale global numerical simulations \citep{Miesch:2008}. We suspect
that this arises because the observations under-resolve the narrow downflow lanes.
Interestingly, the PDFs do not exhibit the same asymmetry within magnetized
regions. In fact, the divergence distribution is rather symmetric, showing
no preference for either convergence or divergence. The magnetic activity
has changed the fundamental nature of the convection. Perhaps magnetoconvection
does not result in the same organized cellular structure that is so apparent
in both observations and simulations of quiet sun. This is certainly the
impression one receives when one examines our flow maps in regions of activity;
however, we have yet to quantitize this property in a meaningful way.

\subsection{Production of Magnetic Shear}
\label{subsec:Shear}

We measure cyclonic rotation about the active region at the periphery of
the active region and anticyclonic motion within the cores of active regions.
Clearly such differential motion results in shear that imparts twist to the
magnetic field within active regions. Let us first consider the cyclonic inflow.
In an active region with a diameter of 300 Mm, a flow of 5 m s$^{-1}$ around
the periphery results in a circumnavigation time for a fluid parcel of roughly
2000 days. This is clearly a very slow windup that is unlikely to result in
significant magnetic shear over the lifetime of the active region. The anticylonic
moat flows from sunspots, on the other hand, are more significant. Assuming
that the moat flow extends out to a radius of 20 Mm with a rotational speed
of 10 m s$^{-1}$ (see Figure~\ref{fig:ARflows}),
a complete winding of the field would occur in 145 days. Thus over the
course of a single Carrington rotation, significant shear can be introduced
into an otherwise stable magnetic configuration. Of course shear of this
nature may aid in the destabilization of the magnetic field, leading to
flares and coronal mass ejections. We suspect that the correlations between
flaring activity and the vorticity measured in low-resolution helioseismic
measurements \citep{Mason:2006} may result from the large-scale shearing motions
observed here.

We note that the measurement of systematic trends in the behavior of the flow
vorticity is difficult since the trends are weak; averaging over a substantial
number of active regions is required. The inflows and outflows, on the other
hand, are quite robust signatures that are easily observed even within a
single active region. This difference arises primarily because the circulations
have an amplitude 4--5 times smaller than the inflows and outflows. Since we
suspect that the systematic cyclonic and anticyclonic motions that produce
magnetic shear are caused by Coriolis deflection of the inflows and outflows,
there is a strong possibility that the inflows and outflows may be a better
predictor of flare activity, simply because of their greater amplitude.

\subsection{Surface Inflows and Flux Confinement}
\label{subsec:Confinement}
	
The surface inflows may play a very dramatic role in the evolution of active
regions. \cite{Hurlburt:2008} have suggested that the observed surface
inflows may inhibit the diffusion of magnetic flux out of active regions,
thereby prolonging the lifetime of an active region before it breaks up and
disperses. By their estimates, the advection of field due to an inflow with
a speed between 10 and 100 m s$^{-1}$ should be sufficient to balance the
outward transport of magnetic field by turbulent diffusion. This is exactly
the range in which our measured inflows fall. Therefore, flows on all scales
may be crucial in the decay of active complexes. Turbulent diffusion of the
field by supergranulation and smaller-scale flows works to disperse the field,
but is counteracted, at least partially, by coalescence of field arising from
inward advection by flows with spatial scales larger than supergranulation.
This argument is predicated on the assumption that the magnetic field is not
structured on the scale of the larger-scale flows. If, for example, the magnetic
flux is concentrated at the boundaries of giant cells, as it is for granules
and supergranules, spatial correlations between the magnetic flux and large-scale
flows would prevent the mean magnetic advection rate from simply being the
product of the mean inflow speed and the mean flux density. However, we have
tested this possibility and found it not to be the case. We compute the mean
field advection rate by the relation

\begin{eqnarray}
	v_{\rm mag} &\equiv& \sum_i  \frac{w_i(B)}{L_i(B) \; \bar{\Phi}_i(B)} \;
		\int_{C_i(B)} |\bvec{\Phi}| \; \bvec{v} \cdot \bvec{\hat{n}} \; dl \; ,	\\
\nonumber \\	
	\bar{\Phi}_i(B) &\equiv& \frac{1}{L_i(B)} \int_{C_i(B)} |\bvec{\Phi}| \; dl \; ,
\end{eqnarray}

\noindent where $|\bvec{\Phi}|$ is the magnetic flux density and $\bar{\Phi}_i(B)$
is the mean field strength along a contour. Note that $B$ is the magnetic
field strength associated with the contour, which is derived from the smoothed
magnetograms; whereas $|\Phi|$ is the field strength within the full-resolution
magnetograms.  We find that, within the observational errors, $v_{\rm mag}$ is identical
to the mean inflow rate shown in Figure~\ref{fig:ARflows}. Thus, the transport
of the magnetic field is dominated by advection by the large-scale flow
component. The consolidation of the field within active regions is therefore
an important mechanism that impedes the dissolution of active regions through
turbulent diffusion. In fact, the inward advection may be sufficiently strong
that flux sequestration becomes a difficulty for models of the evolution of the
sun's global magnetic field, which rely on the steady diffusion of flux from active
regions \citep{DeRosa:2006}. However, before we declare that the surface inflows
problematically inhibit the dispersal of magnetic flux from active regions, an
additional question must be answered. We must ascertain whether the surface
inflows vary over the lifespan of an active region. It is likely that young
active regions possess strong inflows and hold onto their magnetic flux rather
tightly; whereas older regions have weakened inflows, thus enabling the escape
of large amounts of magnetic flux. To date such studies have yet to be performed.

\section{Conclusions}
\label{sec:Conclusions}

Using ring analysis, we have measured the flow field within the upper
2 Mm of the solar convection zone with a spatial resolution of 2$^\circ$.
From these measurements we have found that solar active regions have a
bulk motion with respect to quiet sun as well as large-scale circulation
cells associated with their presence. The bulk motion consists of a 20 m s$^{-1}$
prograde motion relative to quiet sun at the same latitude, with simultaneous
meridional advection that moves in lock step with the surrounding
field-free plasma.

The large-scale circulations that are established resemble, in many ways,
an inverted hurricane (see the schematic diagrams shown as Figures~\ref{fig:ARsideview}
and \ref{fig:ARtopview}). In a hurricane, evaporation from a warm sea
surface drives an upflow and a concomitant inflow to feed the upwelling.
The upflow eventually spreads high above the surface forming an outflow.
Coriolis forces act upon the surface inflows and spin up the storm until
a quasigeostrophic balance is achieved. For active regions, surface cooling
by enhanced radiative losses in plage plays the role of a warm sea surface.
However, instead of causing a warm upwelling, this cooling causes a descending
downdraft. This downdraft pulls in fluid from the surroundings causing an
inflow and the attendent cyclonic motion.

The existence of sunspots modifies this picture somewhat. It may be that
the surface cooling is weaker within sunspots and that sunspots represent
a hole in the downdraft caused by plage. The outflow from the sunspots
could simply be drawn to feed the annular downdraft. However, a more
likely explanation is that the outflows result from a different mechanism
entirely, one that depends crucially on the orientation of the penumbral
filaments. Whatever the source, Coriolis forces acting on these outflows
produce a measurable flow deflection, resulting in a net anticyclonic rotation
about the sunspot. While this rotation is fairly weak, over the lifetime
of an active region, the rotation should produce significant shear, perhaps
playing a role in destabilizing the coronal magnetic field overlying the
active region.

Superimposed on these circulations are convective flows. We find that quiet
sun exhibits a notable asymmetry where a larger percentage of the solar surface
is covered with outflows than inflows. For the spatial scales that our
helioseismic technique samples, we find that within active regions this
asymmetry disappears, and parity between convergence and divergence is attained.
Why this occurs isn't obvious; however, it may have something to do with the
disruption or segmentation of the larger convective scales by the presence of
magnetism. Our measurement technique is only capable of measuring flows with a
spatial scale larger than 2$^\circ$ in heliographic angle ($\approx$ 24 Mm).
Therefore, we sample flows larger than supergranulation.  It would be quite
useful to perform a similar study with finer resolution where supergranules
are explicitly resolved.


\acknowledgments

We are indebted to to Richard S. Bogart for his substantial efforts in
tracking and remapping the MDI data for use in ring analyses. We gratefully
thank Greg Kuebler for using his artistic talents to produce Figures~\ref{fig:ARsideview}
and \ref{fig:ARtopview}. We acknowledge support from NASA through grants
NAG5-13520, NNG05GM83G, NNG06GD97G, NNX07AH82G, NNX08AJ08G, and NNX08AQ28G.




\newcommand{\jaa}{{\sl J. Astrophys. Astron.}}

\def\figone{%
\begin{figure*}%
        \epsscale{1.0}%
        \plotone{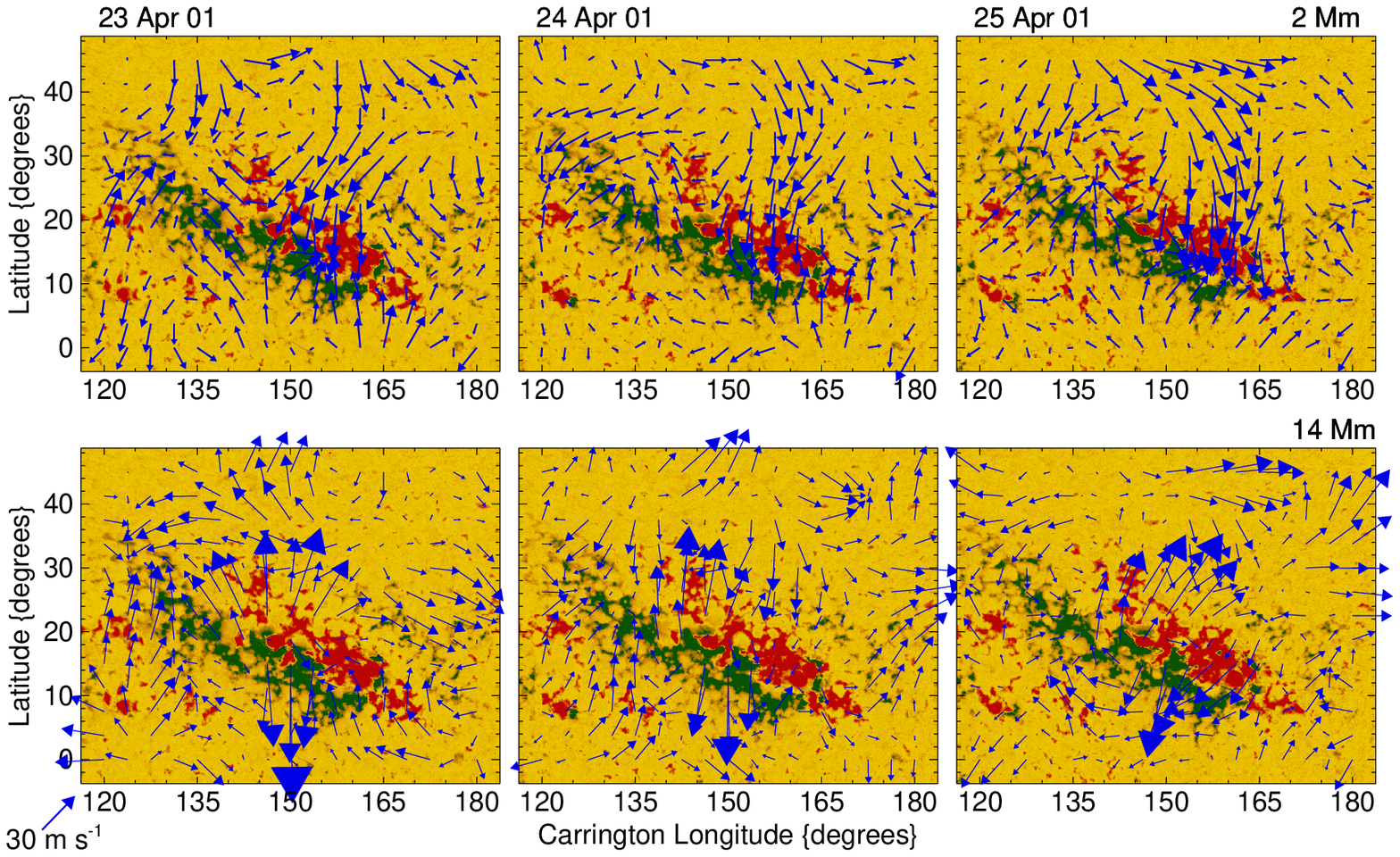}%
        \caption{\small Flows obtained by dense-pack ring analysis near
NOAA AR9433, a large active complex which underwent multiple flux emergence
events and survived several disk passages. The dense-pack procedure is a
technique that samples deep flows but with a fairly low resolution of
15$^\circ$. The flow field (blue arrows) over three consecutive days is
shown at two different depths (2 Mm and 14 Mm). Underlying the flow arrows
are magnetograms with red and green tones indicating opposite polarity.
The flows converge on the active region near the surface, while at depth
the active region is a strong outflow site. Adapted from \cite{Haber:2004}.
\label{fig:DensePack} }%

\end{figure*}%
}


\def\figtwo{%
\begin{figure*}%
        \epsscale{0.5}%
        \plotone{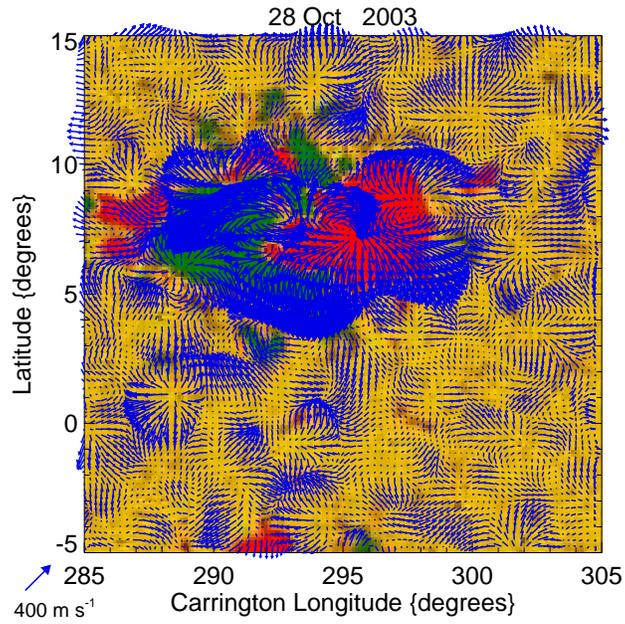}%
        \caption{\small Flows obtained using
HRRA around a newly emerged active region. The flows were inferred from
MDI data obtained on 28 October 2003. The region first appeared two days
previously and at the time of the helioseismic observations still possessed
a compact magnetic field distribution with little outlying plage, despite
the continued emergence of multiple dipole structures. The flows around
this young active region are predominantly surface outflows originating
from the sunspots.
\label{fig:EmergingAR} }%

\end{figure*}%
}


\def\figthree{%
\begin{figure*}%
        \epsscale{1.0}%
        \plotone{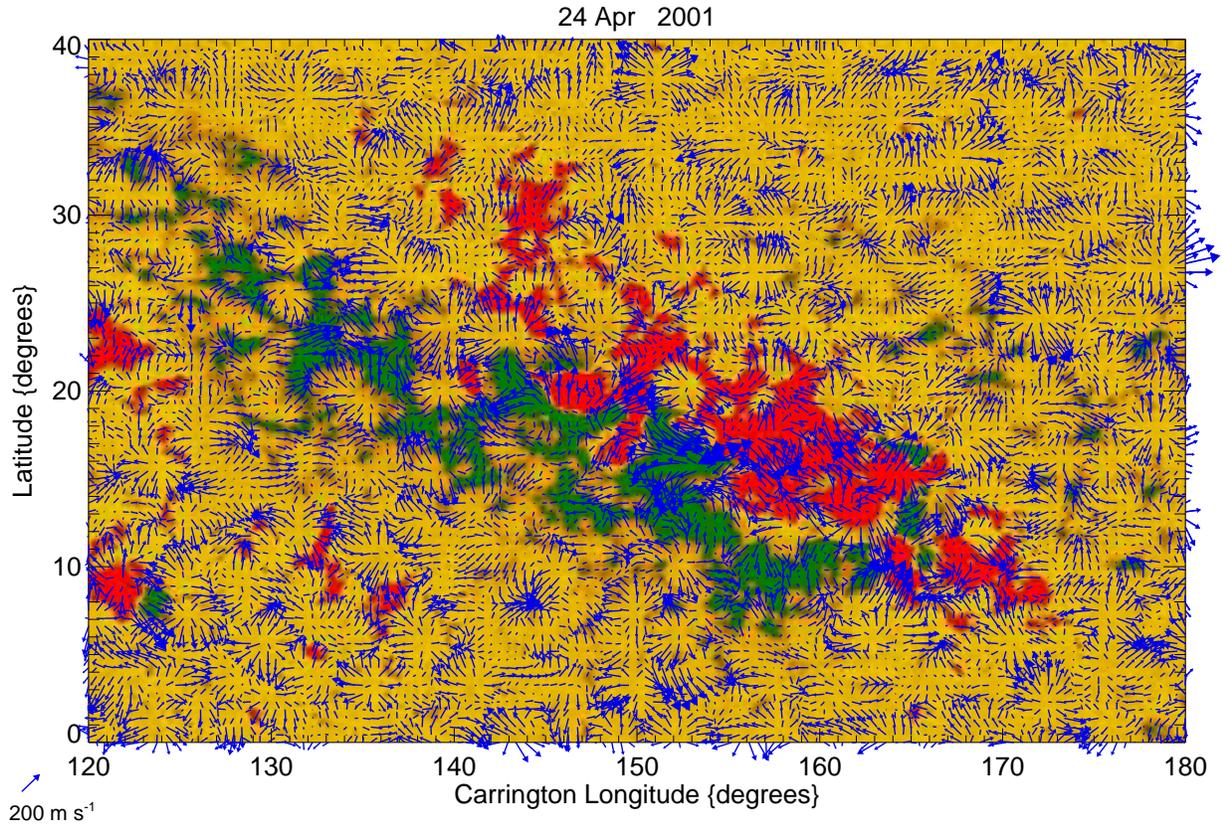}%
        \caption{\small Flows obtained using HRRA in the vicinity of
a mature active complex, NOAA AR9433. The flows were inferred from MDI
data obtained on 24 April 2001. The regular cellular pattern is due to
solar convection with a spatial scale slightly larger than supergranulation.
The cell boundaries often coincide with concentrations of magnetic flux,
but within regions of strong magnetic field the flow structure becomes
complex and less ordered.
\label{fig:LargeAR} }%

\end{figure*}%
}


\def\figfour{%
\begin{figure*}%
        \epsscale{1.0}%
        \plotone{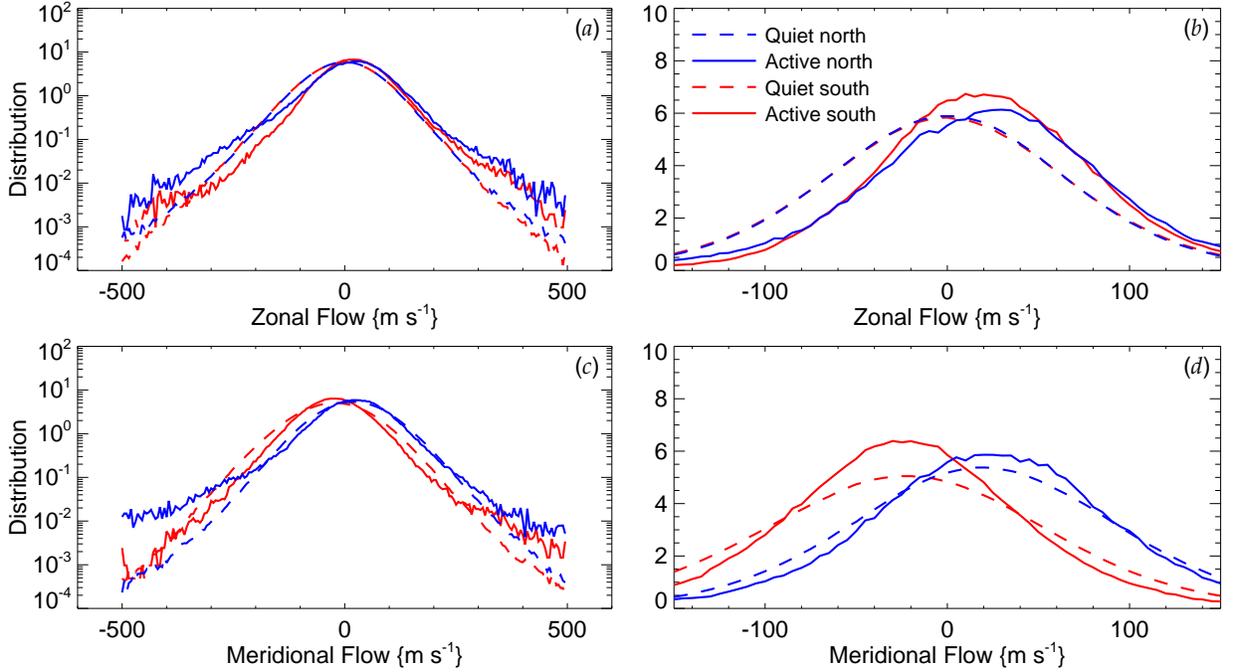}%
        \caption{\small Probability density functions for the zonal
flow (top panels) and for the meridional flow (bottom panels). The left
panels are logarithmically scaled to enhance the wings of the distributions,
whereas the right panels are linearly scaled to enhance the core of the
distributions. The solid curves correspond to distributions for magnetized
pixels ($B>50$ G) and the dashed curves to quiet pixels. The blue (red)
curves show the distributions for only the regions that lie within a
$10^\circ$-wide latitudinal band centered at $30^\circ$ north (south) of
the equator. From the peaks of the distributions, it is clear that magnetized
pixels rotate more rapidly than quiet regions (by about 20 m s$^{-1}$) while
those same regions are advected poleward at roughly the same rate as quiet
regions.
\label{fig:ZonMerPDFs} }%

\end{figure*}%
}


\def\figfive{%
\begin{figure*}%
        \epsscale{1.0}%
        \plotone{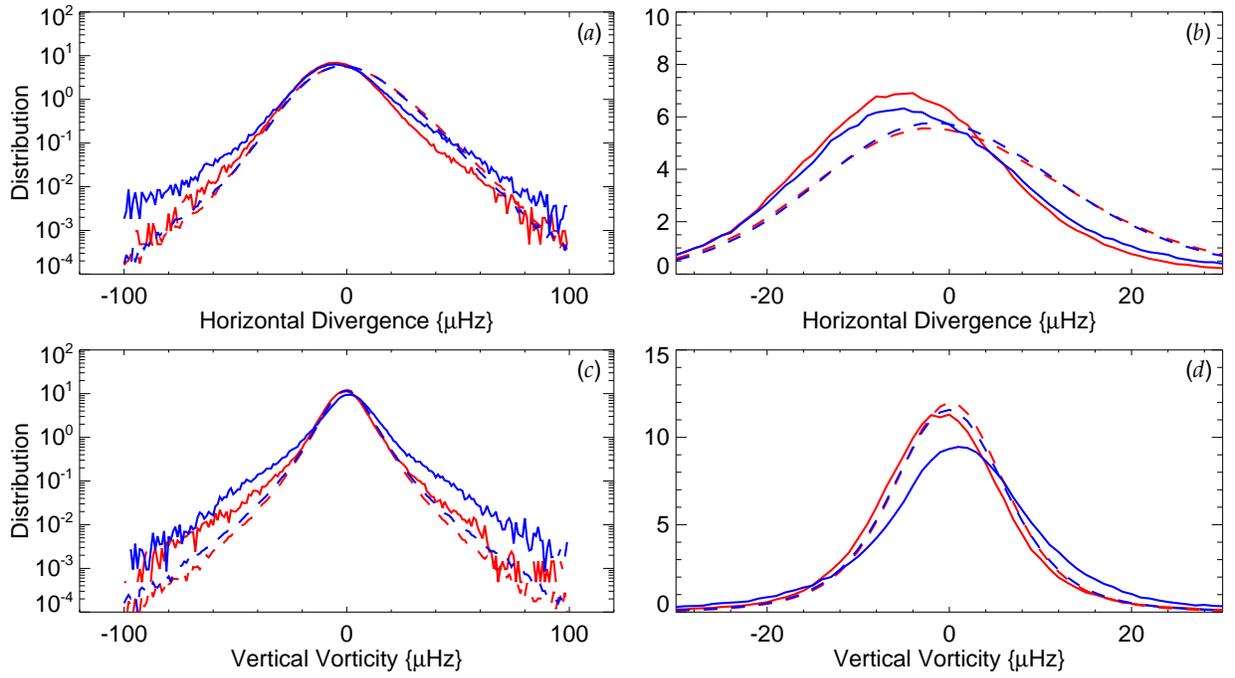}%
        \caption{\small The same as Fig.~\ref{fig:ZonMerPDFs} except that the
distributions are for the horizontal divergence (top panels) and the vertical
component of the vorticity (bottom panels). Magnetized pixels are preferentially
zones of convergence. There is also a weak tendency for cyclonic vortical
motion in magnetic regions.
\label{fig:DivCurlPDFs} }%

\end{figure*}%
}


\def\figsix{%
\begin{figure*}%
        \epsscale{1.0}%
        \plotone{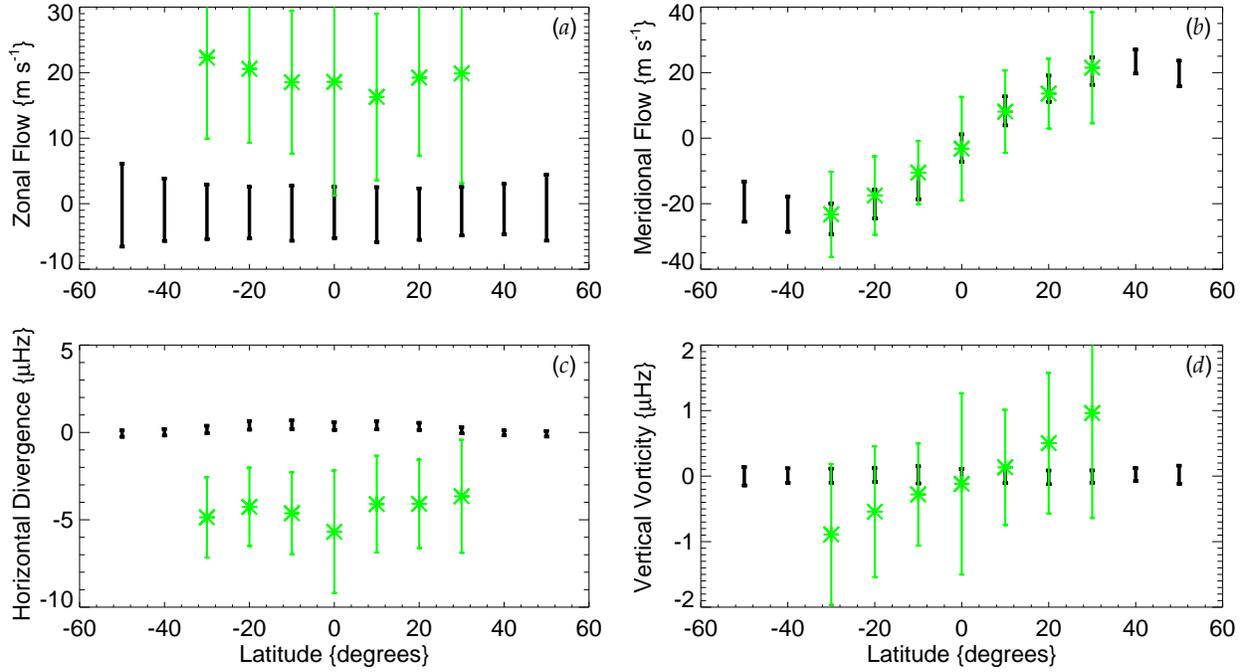}%
       \caption{\small Mean values of the distribution functions as a
function of latitude. The thick black bars correspond to the quiet pixels
and the green bars to the magnetized pixels. The length of the bars indicates
a typical variance between means computed over different days of data. ($a$)
Mean of the zonal flow distribution showing that magnetized regions rotate
more quickly than quiet regions by roughly 20 m s$^{-1}$. ($b$) Mean of the
meridional flow distributions showing that active and quiet regions move
poleward at the same speed. ($c$) Mean of the divergence distribution showing
that magnetized regions are zones of convergence, independent of latitude.
($d$) Mean of the vorticity distribution demonstrating that magnetized
pixels are zones of cyclonic vorticity.
\label {fig:MeanFlows} }%

\end{figure*}%
}


\def\figseven{%
\begin{figure*}%
        \epsscale{1.0}%
        \plotone{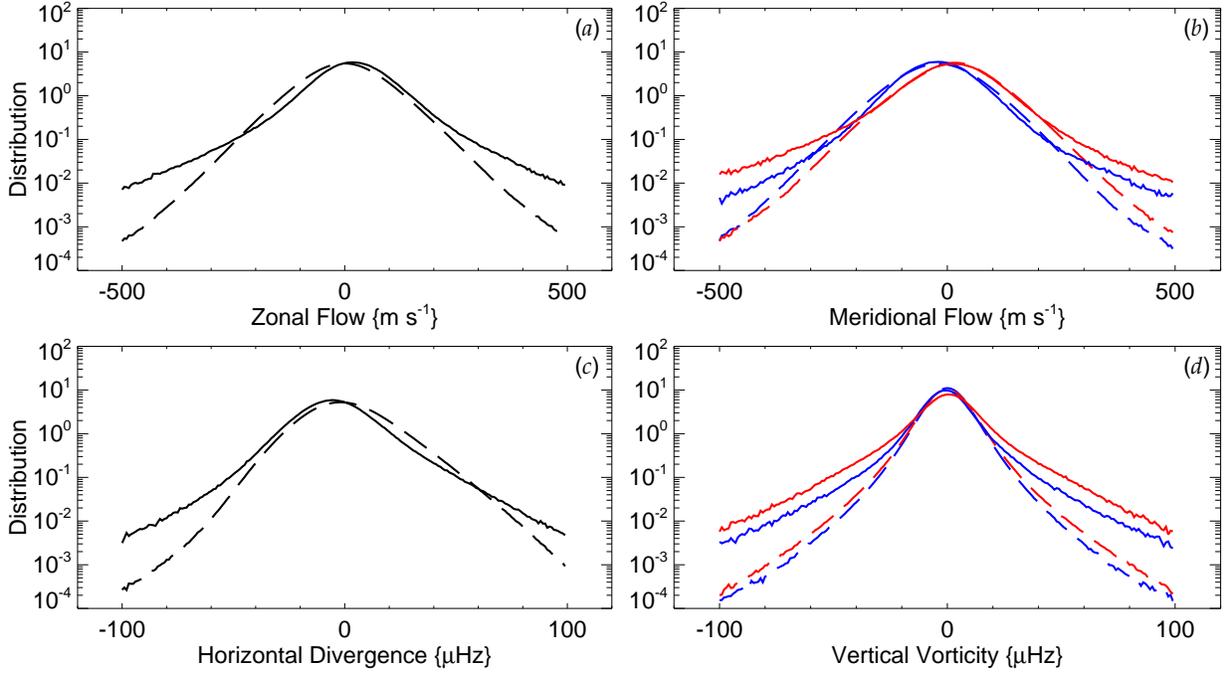}%
        \caption{\small Flow distributions averaged over latitudinal bands.
The solid curve corresponds to magnetized regions and the dashed to quiet
regions. ($a$) Zonal flow distributions averaged over all latitudes. ($b$)
Meridional flow distributions averaged over each hemisphere (red--southern
and blue--northern). ($c$) Distributions of the horizontal divergence averaged
over all latitudes. ($d$) Distributions of the vertical component of the vorticity
averaged over each hemisphere. From the shapes of these distributions, one
may deduce that the flow fields within regions of magnetic activity have an
expanded range of speeds. Furthermore, in quiet sun, the divergence distribution
indicates that more of the solar surface is covered with outflows than inflows,
whereas within magnetized regions outflows and inflows have an equal filling
factor.
\label {fig:MeanPDFs} }%

\end{figure*}%
}


\def\figeight{%
\begin{figure*}%
        \epsscale{0.5}%
        \plotone{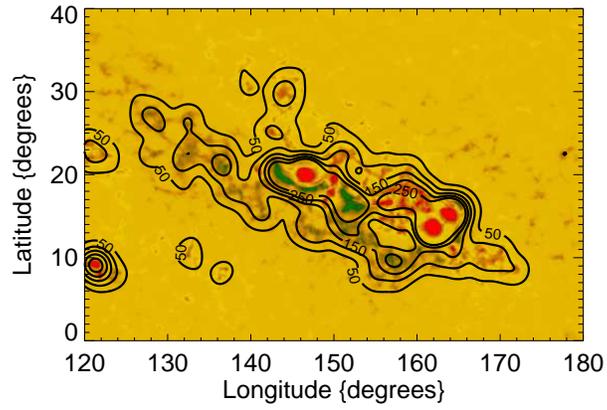}%
        \caption{\small MDI magnetogram of NOAA AR9433. The contours shown
were computed from a smoothed map of the unsigned magnetic flux density.
The strength of the contour is measured in Gauss.
\label {fig:MagGram} }%

\end{figure*}%
}


\def\fignine{%
\begin{figure*}%
        \epsscale{0.5}%
        \plotone{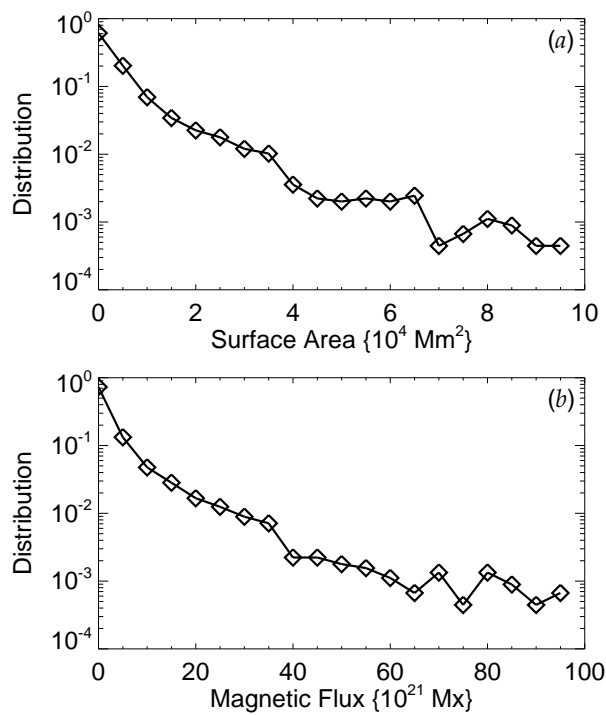}%
        \caption{\small Distribution of ($a$) areas and ($b$) magnetic flux
for all flux concentrations identified during three periods of time:
1 March--26 May 2001, 11 January--21 May 2002 and 12 September--15 November 2003.
These flux concentrations were identified from daily MDI magnetogram images
and all lie within $60^\circ$ of disk center. Our sample of active regions
is composed of those with an area exceeding $1\times10^4$ Mm$^2$. In total
661 flux concentrations were identified, although many of these are the same
flux concentrations measured on subsequent days.  Taking this into account,
roughly 100 independent flux concentrations compose the sample of active regions.
\label {fig:DistARs} }%

\end{figure*}%
}


\def\figten{%
\begin{figure*}%
        \epsscale{0.5}%
        \plotone{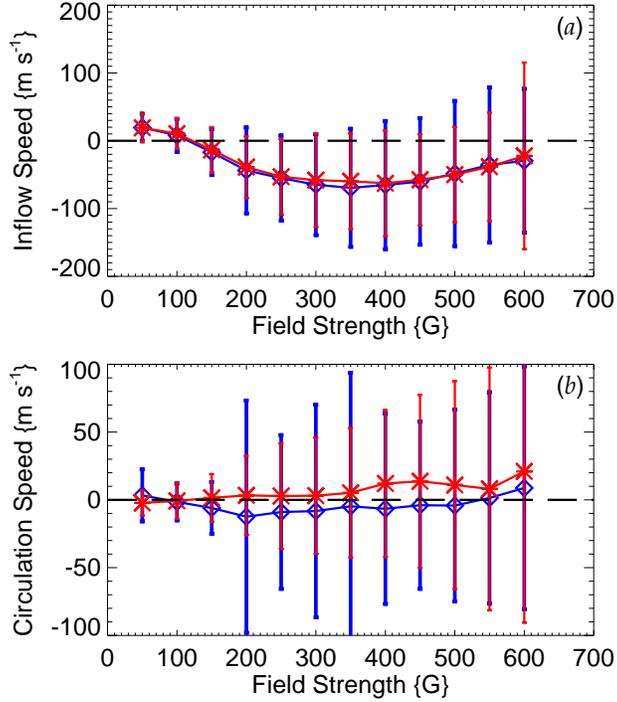}%
        \caption{\small ($a$) Inflow speeds at different magnetic contour
levels within active regions. The blue curve shows the average over all
active regions located in the northern hemisphere and the red shows the
southern hemisphere. The bars indicate the variance for any particular
active region about the mean. The periphery of active regions possess positive
inflows with an amplitude of 20--30 m s$^{-1}$. As the field strength
increases towards the interior of active regions, this inflow turns into
an outflow presumably forming a downflow where the flows collide. The very
core of active regions, where the sunspots reside, are zones of strong outflow
($\approx$ 50 m s$^{-1}$). ($b$) Circulation speeds around the magnetic
contours in a counterclockwise direction. There is a tendency for active
regions to possess cyclonic motion at their peripheries and anticyclonic
motion in their cores.
\label {fig:ARflows} }%

\end{figure*}%
}


\def\figeleven{%
\begin{figure*}%
        \epsscale{1.0}%
        \plotone{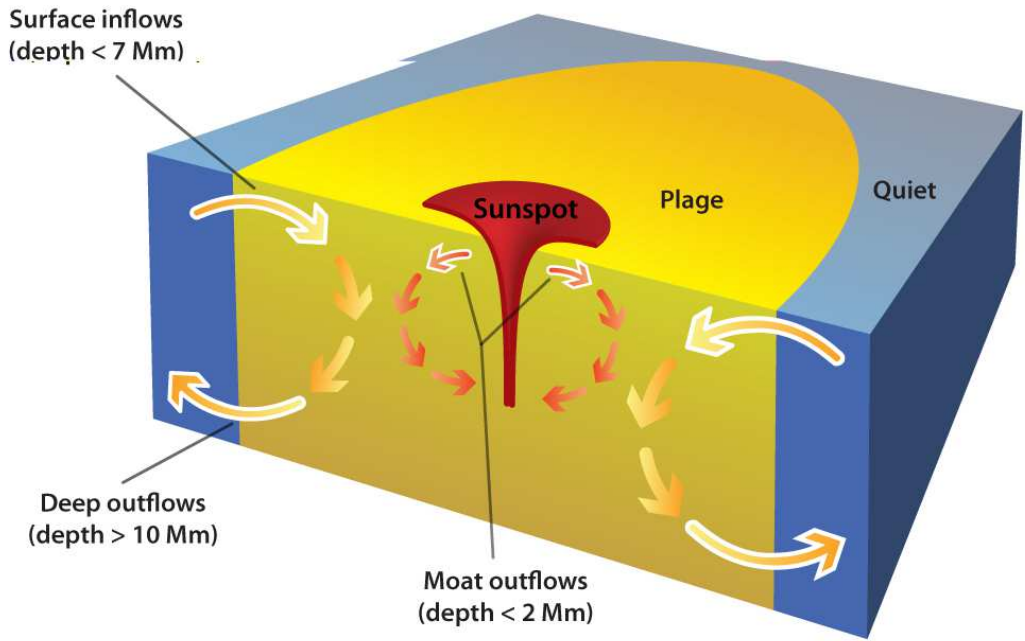}%
        \caption{\small Schematic diagram providing a sideview of the large-scale
circulations established within active regions. Surface cooling within
the plage results in a downdraft which draws fluid in at the surface. Therefore,
there is a mean inflow ($\approx$ 20 m s$^{-1}$) at the active region's periphery.
Below a depth of 10 Mm, an outflow manifests that is likely the return flow
connected to the surface inflow. The moat flows streaming out from sunspots
at the surface impact this inflow somewhere within the plage and presumably
join in the down flow. The arrows outlined in white have been observed
through a variety of techniques, including local helioseismology and correlation
tracking.
\label {fig:ARsideview} }%

\end{figure*}%
}


\def\figtwelve{%
\begin{figure*}%
        \epsscale{1.0}%
        \plotone{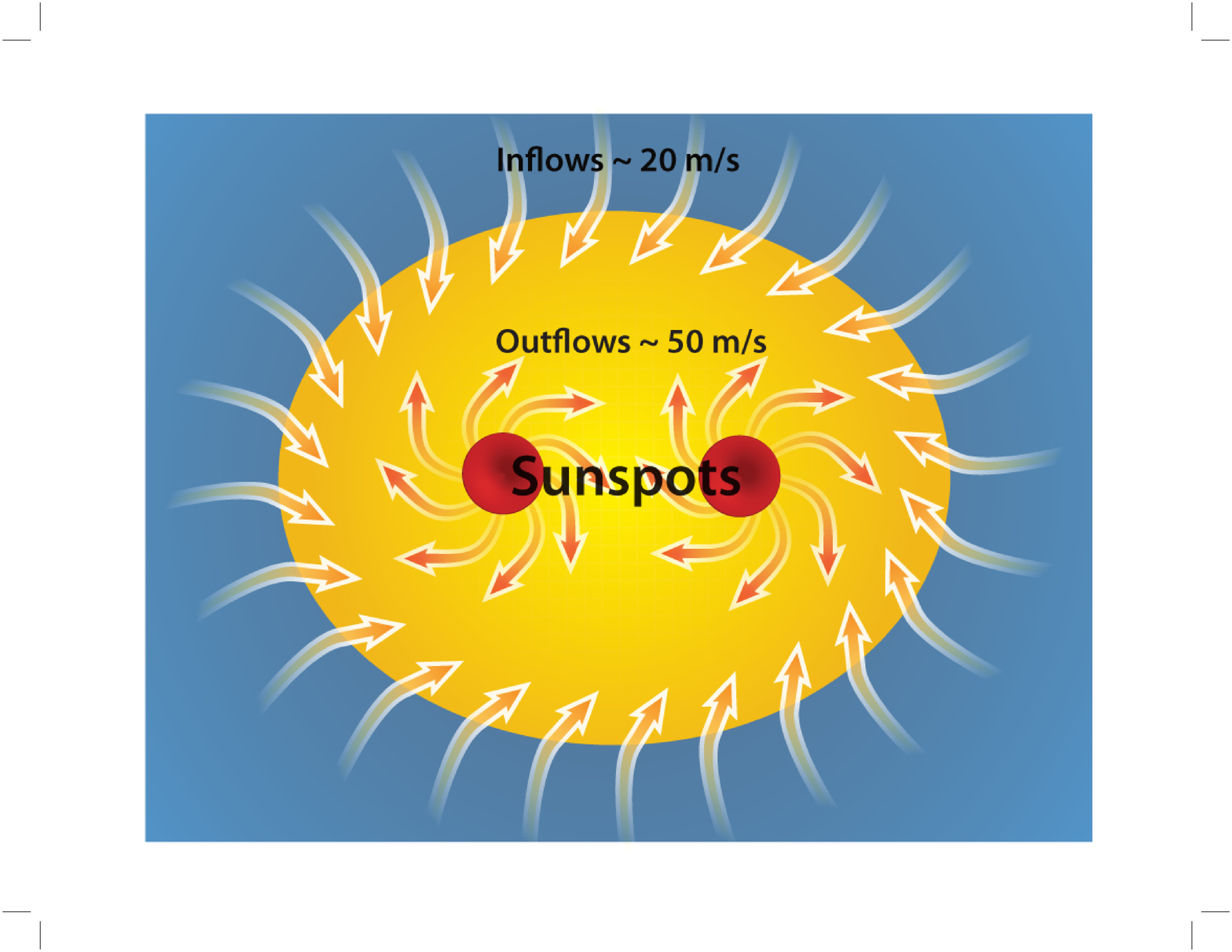}%
        \caption{\small Schematic diagram of the surface flows that form
within active regions in the northern hemisphere. The inflow into the active region and the outflow
from sunspots are both deflected by Coriolis forces producing cyclonic
circulations at the outer boundary of the active region and anticyclones
at the site of the sunspots.
\label {fig:ARtopview} }%

\end{figure*}%
}

\figone
\figtwo
\figthree
\figfour
\figfive
\figsix
\figseven
\figeight
\fignine
\figten
\figeleven
\figtwelve

\end{document}